\documentstyle[aps,preprint,prc,12pt]{revtex}

\begin{document}

\title{\bf Microscopic nuclear equation of state with three-body 
        forces and neutron star structure}

\author{M. Baldo, G.F. Burgio }

\address{ \it Dipartimento di Fisica, Universit\'a
 di Catania and I.N.F.N. Sezione di Catania, c.so Italia 57, 
I-95129 Catania, Italy}

\author{and I. Bombaci}

\address{ \it Dipartimento di Fisica, Universit\'a di Pisa and 
          I.N.F.N. Sezione di Pisa, Piazza Torricelli 2, 
I-56100 Pisa, Italy}

\maketitle

\bigskip


\bigskip

\newpage

\begin{abstract}
We calculate static properties of non-rotating neutron stars (NS's) using  
a microscopic equation of state (EOS) for asymmetric nuclear matter.  The
EOS is computed in the framework  of the Brueckner--Bethe--Goldstone 
many--body theory.   We introduce three-body forces in order to reproduce
the correct saturation  point of nuclear matter. A microscopic well behaved
EOS is derived. We obtain a maximum mass configuration with $M_{max} = 1.8
M_\odot$,   a radius $R = 9.7$ km and a central density $n_c =
1.34~fm^{-3}$.   We find the proton fraction exceeds the critical value
$x^{Urca}$,  for the onset of direct Urca processes, at densities  $n \geq
0.45~fm^{-3}$.   Therefore, in our model, NS's with masses above $M^{Urca}
= 0.96 M_\odot$   can undergo very rapid cooling depending on whether or
not nucleon  superfluidity in the interior of the NS takes place.   A
comparison with other microscopic models for the EOS is done, and neutron 
star structure is calculated for these models too.  

\end{abstract}
\bigskip
\bigskip

{PACS numbers: 97.60.Jd, 21.65.+f }

\bigskip

\newpage

In the next few years it is expected that a large amount of novel 
informations on neutron stars (NS's) will be available from the  new
generation of X--ray and $\gamma$--ray satellites.  Therefore, a great
interest is devoted presently to the study of  NS's and to the prediction
of their structure on the basis of the  properties of dense matter.  The
equation of state (EOS) of NS matter covers a wide density range,  from
$\sim 10$ g/cm$^3$ in the surface to several times nuclear matter 
saturation density ($\rho_0 \sim 2.8~10^{14}$ g/cm$^3$) in the center  of
the star\cite{sha}.  The interior part (core) of a NS is made by asymmetric
nuclear matter  with a certain lepton fraction.   At ultra--high density, 
matter might suffer a transition to other exotic hadronic components (like
hyperons, a $K^-$ condensate or a  deconfined phase of quark matter). The
possible appearance of such an  exotic core has enormous  consequences for
the neutron star and black hole  formation mechanism\cite{bomb}.    
Unfortunately large uncertainities are still present in the theoretical 
treatment of this ultra--dense regime\cite{gle,pra+}.  Therefore, in the
present work, we consider a more conventional picture  assuming the NS core
is composed only by an uncharged mixture of neutrons,  protons, electrons
and muons in equilibrium with respect to the  weak interaction 
($\beta$--stable matter).   Even in this picture, the determination of the
EOS of asymmetric nuclear  matter to describe the core of the NS, remains a 
formidable theoretical  problem\cite{mart}.  

Any ``realistic'' EOS must satisfy several requirements : i) It must
display the correct saturation point for symmetric nuclear matter (SNM);
ii) it must give a symmetry energy  compatible with nuclear phenomenology
and well behaved at high densities; iii) for SNM the incompressibility at
saturation must be compatible with phenomenology on monopole nuclear
oscillations\cite{swi};  iv) both for neutron matter (NEM) and SNM the
speed of sound must not exceed the speed of light (causality condition), at
least up to the relevant densities; the latter condition is automatically
satisfied only in fully relativistic theory.

In this letter we present results for some NS properties obtained on the 
basis of a microscopic EOS, recently developed\cite{tbf}, which satisfies 
requirements i-iv, and compare them with the predictions of other 
microscopic EOS's.  The Brueckner-Hartree-Fock (BHF) approximation for the
EOS in SNM, within the  continuous choice \cite{bhf}, reproduces closely
the  Brueckner--Bethe--Goldstone  (BBG) results which include up to four
hole line diagram  contributions\cite{Day}, as well as the variational
calculations \cite{var},  at least up to few times the saturation density. 
Non--relativistic calculations, based on purely two--body interactions, 
fail to reproduce the correct saturation point for SNM. This well known 
deficiency is commonly corrected introducing three-body forces (TBF). 
Unfortunately, it seems not possible to reproduce the experimental binding 
energies of light nuclei and the correct saturation point accurately with
one   simple set of  TBF \cite{var}.  Relevant progress has been made in
the theory of nucleon TBF, but a complete theory is not yet available.  In
ref.\cite{var} a set of simple TBF  has been introduced within the  
variational approach. We introduced \cite{tbf} similar TBF  within the BHF 
approach, and we have adjusted the parameters in order to reproduce closely
the correct saturation point of SNM, since for NS studies this is an
essential requirement, and there is no reason to believe that TBF be the
same as in light nuclei.   The corresponding EOS (termed BHF3) is depicted
in Fig. 1, in comparison  with the EOS obtained in BHF approximation
without three-body forces (BHF2),  but using the same two-body force, i.e.
the Argonne $v_{14}$ ($Av_{14}$)  potential \cite{arg}).   In the same
figure, we show the variational EOS (WFF) of ref.\cite{var}  for the 
$Av_{14} + TBF$ Hamiltonian, and the EOS from a recent  Dirac-Brueckner
calculation (DBHF) \cite{dbhf} with the Bonn--A two--body  force.       The
BHF3 EOS saturates at  $n_o = 0.18~fm^{-3},  E = -15.88~MeV$, and is
characterized by an incompressibility $K_{\infty} = 240~MeV$,   very close
to the recent phenomenological estimate of ref.\cite{swi}.  In the low
density region ($n < 0.4~fm^{-3}$), BHF3 and DBHF equations of  state are
very similar. At higher density, however, the DBHF is stiffer  than the
BHF3.   The discrepancy between these two models for the EOS can be easily
understood by noticing that the DBHF treatment is equivalent \cite{tao} to
introduce in the non-relativistic BHF2 the three-body force corresponding
to the excitation of a nucleon-antinucleon pair, the so-called
Z-diagram\cite{zeta}.  The latter is repulsive at all densities. In BHF3
treatment, on the contrary, both attractive and  repulsive three-body
forces are introduced \cite{var}, and therefore a softer EOS can be
expected.  

\noindent
Fractional polynomial fits to each one of these EOS's allow to compute the
corresponding pressure and speed of sound $c_s$ to compare with the  speed
of light $c$. The ratio  $c_s/c$ for all four EOS's as a function of  the
number density is reported in Fig. 2.  WFF model (circles) violates the 
causality condition at  densities  encountered in the core of stars near
the maximum mass configuration for  that model (see fig. 4b) and tab. I).   
The DBHF calculations need an extrapolation to slightly higher densities
than   the largest one considered in ref.\cite{dbhf}. The extrapolation was
done in  such a way to keep the causality condition fulfilled.  The same
procedure was followed for BHF3. In the latter case the BHF procedure was
well converging up to densities $n = 0.76~ fm^{-3}$ for SNM and  $n =
0.912~ fm^{-3}$ for NEM. For DBHF the causality condition was  fulfilled in
the extrapolated region only if particular choices of the fitting  
parameters were used, while for the BHF3 the results were insensitive to a 
wide range of variation of the parameters \cite{bbb}. 

\noindent
It has to be stressed that the $\beta$-stable matter EOS  is strongly
dependent on the nuclear symmetry energy, which in turns  affects the
proton concentration \cite{asym}. The latter quantity is crucial for the
onset of direct Urca processes \cite{urca}, whose occurrence enhances
neutron star cooling rates. In our approach, from the difference of the
energy  per particle $E/A$ in NEM and SNM the symmetry energy $E_{sym}$ can
be extracted assuming a parabolic dependence on the asymmetry parameter
$\beta = {{(n_n - n_p)} / {(n_n + n_p)}}$, being $n_n$ and $n_p$
respectively the neutron and proton number density. This procedure turns
out to be quite reliable\cite{asym}. The values of $E_{sym}$ for the
different EOS's are reported in Fig. 3, together with the corresponding
proton fraction $x = (1 - \beta)/2$. We notice that in both relativistic
and non--relativistic Brueckner--type   calculations, the proton fraction 
can exceed the "critical" value  $x^{Urca} = (11-15)\%$ needed for the
occurrence of direct Urca  processes \cite{urca}.  This is at variance with
the WFF variational calculation (Fig. 3, circles),  which predicts a low
absolute value both for the simmetry energy and  the proton fraction with a
slight bend over.  For BHF3 model we find $x^{Urca} = 13.6\%$, which
correspond to a critical  density $n^{Urca} = 0.447~fm^{-3}$.  Therefore,
BHF3 neutron stars with a central density higher than $n^{Urca}$ develop
inner cores in which direct Urca  processes are allowed.       

The EOS for $\beta$--stable matter  can be used in the 
Tolman--Oppenheimer--Volkoff \cite{tov} equations to compute the neutron
star  mass and radius as a function of the central density.  For the outer
part of the neutron star we have used the equations of state by
Feynman-Metropolis-Teller \cite{fmt} and Baym-Pethick-Sutherland
\cite{bps},  and for the middle-density regime
($0.001~fm^{-3}<n<0.08~fm^{-3}$) we use the  results of Negele and
Vautherin \cite{nv}. In the high-density part  ($n > 0.08~fm^{-3}$) we use
alternatively the three EOS's discussed above.  The results are reported in
Fig. 4.  We display the gravitational mass $M_G$, in units of solar mass 
$M_{\odot}$ ($M_{\odot} = 1.99~10^{33}$ g),  as a function  of the radius R
(panel (a)) and the central number density $n_c$ (panel (b)). As expected,
the stiffest EOS (DBHF) we used in the present  calculation gives higher
maximum mass and lower central density with  respect to the
non-relativistic Brueckner models.    The maximum NS mass for the BHF3 is
intermediate between BHF2 and DBHF, but closer to the latter.  The
difference between BHF3 and WFF neutron stars reflects the discrepancy 
already noticed for the EOS and mainly for the symmetry energy.  This point
will be discussed in more details in a forthcoming paper\cite{bbb}. Table I
summarizes maximum mass configuration properties, for the different  EOS's
used in the present work.

\noindent
In conclusion, we computed some properties of NS's on the basis of a 
microscopic EOS obtained in the framework of BBG many--body theory with 
two-- and three--body nuclear interactions. BHF3 EOS satisfies the general 
physical requirements (points i--iv) discussed in the introduction.  This
is the main feature which distinguishes our BHF3 EOS with respect to other 
microscopic non--relativistic EOS\cite{var,nor}.     The calculated maximum
mass is in agreement with observed NS masses\cite{vanK}.  We found that the
neutron star core is ``proton rich''.     In fact, the proton fraction at
the center of the maximum mass configuration  in the BHF3 model is $x =
26\%$. Our BHF3 neutron stars with mass above the  critical value $M^{Urca}
\equiv M_G(n^{Urca}) = 0.96 M_\odot$  develop   inner cores in which direct
Urca processes can take place.    These stars cool very rapidly or not
depending on the properties of nuclear   superluidity (values of the
superfluid gaps, critical temperatures, density  ranges for the superfluid
transition)\cite{page,bald+}.      Our EOS offers the possibility for a
selfconsistent microscopic calculation for  both the neutron star
structure, and nuclear superfluid properties   within the same many--body
approach and with the same nuclear interaction.

\newpage

\begin{table} 
\begin{center}
\begin{tabular}{|cccc|}
EOS&$M_G/M_{\odot}$&R(km)&$n_c(fm^{-3})$ \\
\hline
DBHF&2.063&10.39          &1.13 \\
BHF3  & 1.794     & 9.74  & 1.34 \\
BHF2  & 1.59      & 7.96  & 1.95\\
WFF   & 2.13      & 9.4   & 1.25 \\
\end{tabular}
\par
\caption{Parameters of the maximum mass configuration: 
the ratio $M_G/M_{\odot}$ is shown for several
EOS's vs. the corresponding radius R and central number density
$n_c$.}
\end{center}
\end{table} 
\newpage
\begin{figure} 
\caption[]
{\footnotesize{The energy per baryon E/A is plotted vs. the 
number density n  for symmetric matter (panel (a)) and 
for neutron matter (panel (b)). 
Several EOS's are shown, {\it i.e.} non-relativistic Brueckner calculations
without (short dashes, BHF2) and with three-body forces (solid line, BHF3)
and a relativistic Dirac-Brueckner one (long dashes, DBHF). 
For comparison a variational calculation 
is also reported (circles, WFF).}}
\end{figure} 
\begin{figure} 
\caption[]
{\footnotesize{The ratio $c_s/c$ is plotted 
as a function of the number density for pure neutron matter.
The different curves refer to BHF2 (short dashes), BHF3 (solid line),
DBHF (long dashes) and the
variational calculations of ref.\cite{var} (circles).}} 
\end{figure} 
\begin{figure}
\caption[]
{\footnotesize{The simmetry energy and the proton fraction
are shown vs. number density respectively in panel (a) and (b).
The different curves refer to BHF2 (short dashes), BHF3 (solid line),
DBHF (long dashes) and the variational calculation WFF (circles).}} 
\end{figure} 
\begin{figure}
\caption[]
{\footnotesize{The gravitational mass $M_G$, expressed in units
of solar mass $M_{\odot}$, is displayed vs. radius R
and the central number density $n_c$. The notation is the same 
as in previous figures.}} 
\end{figure} 
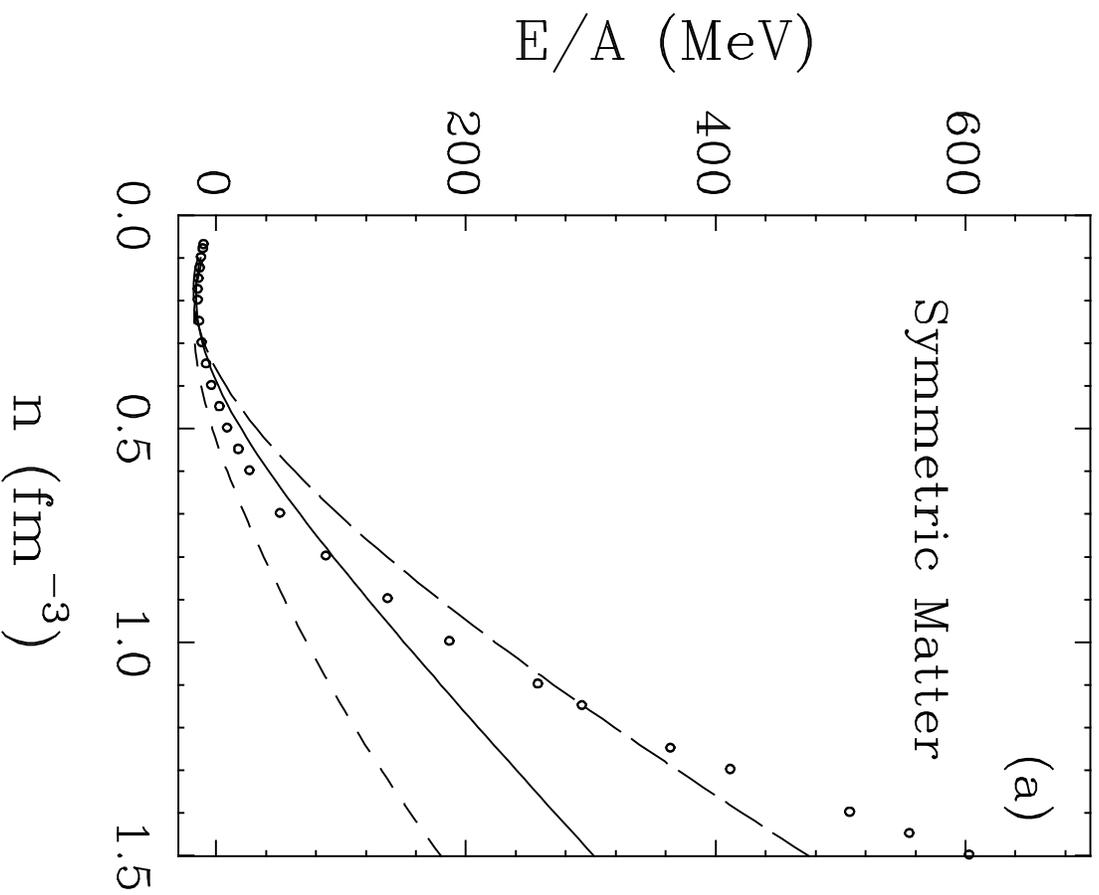
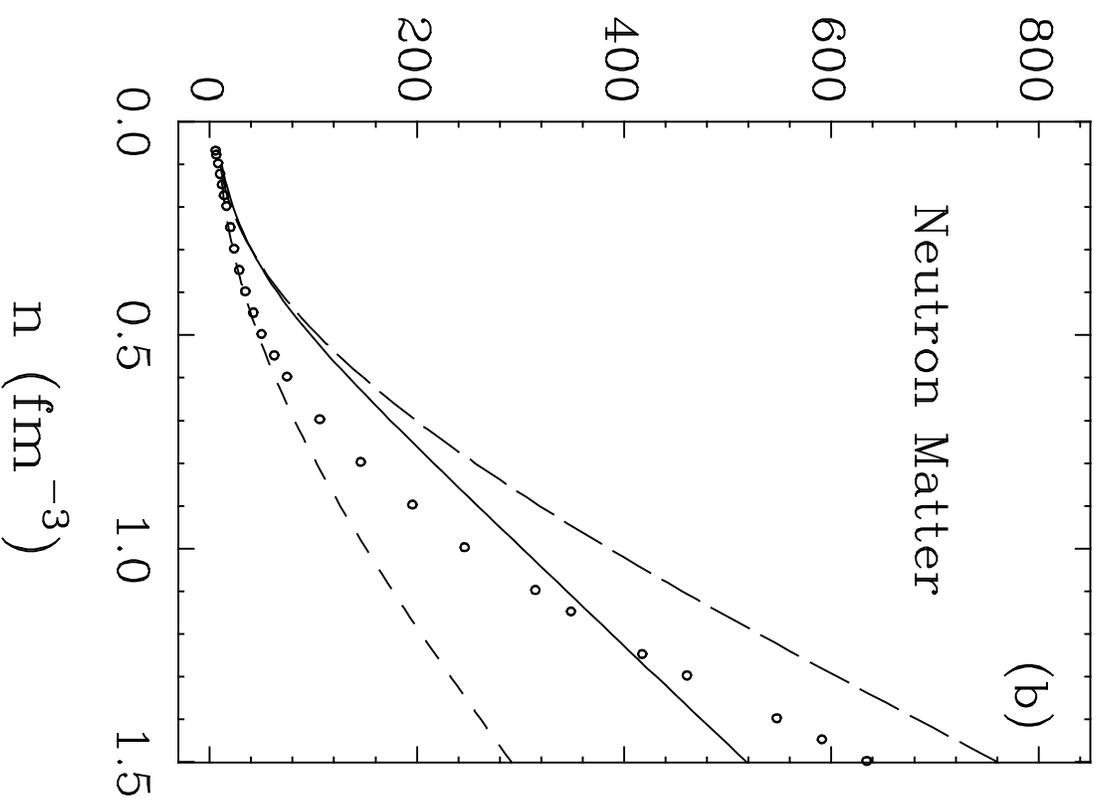
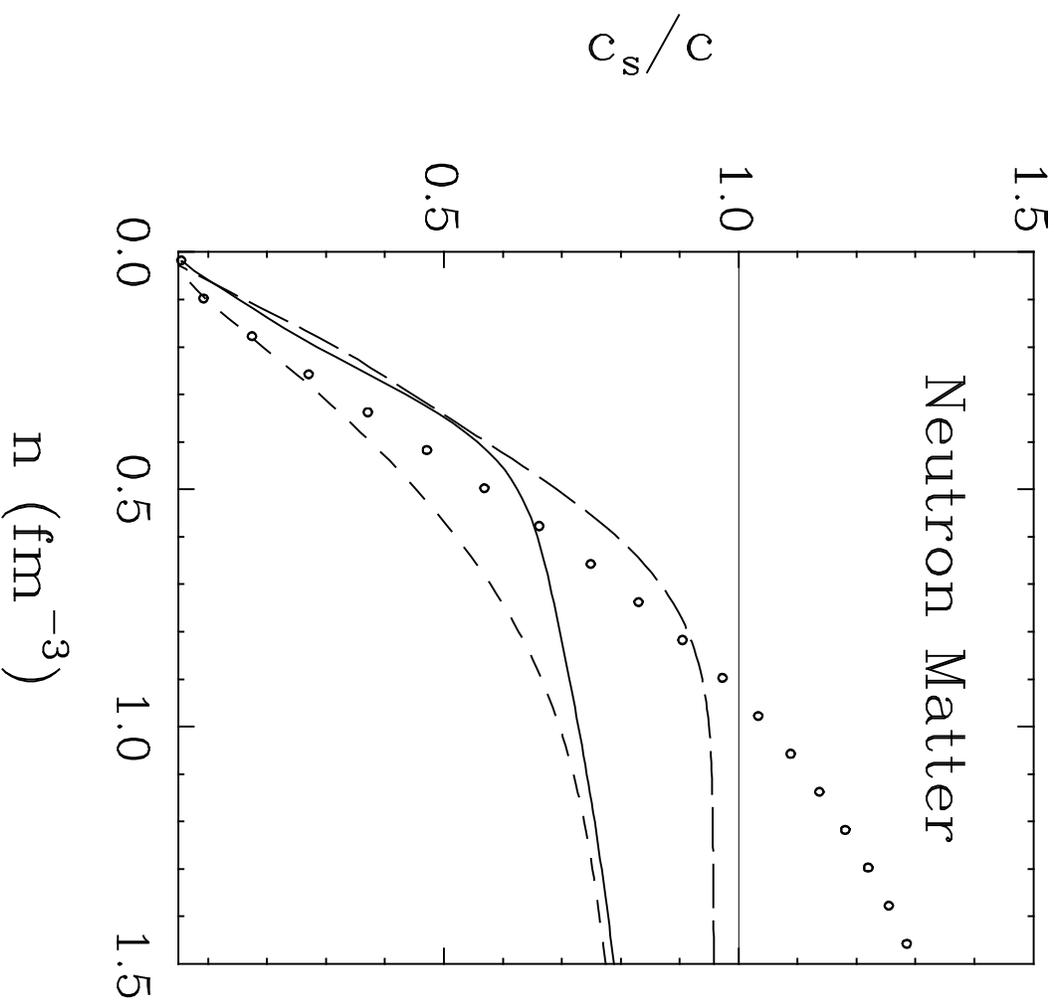
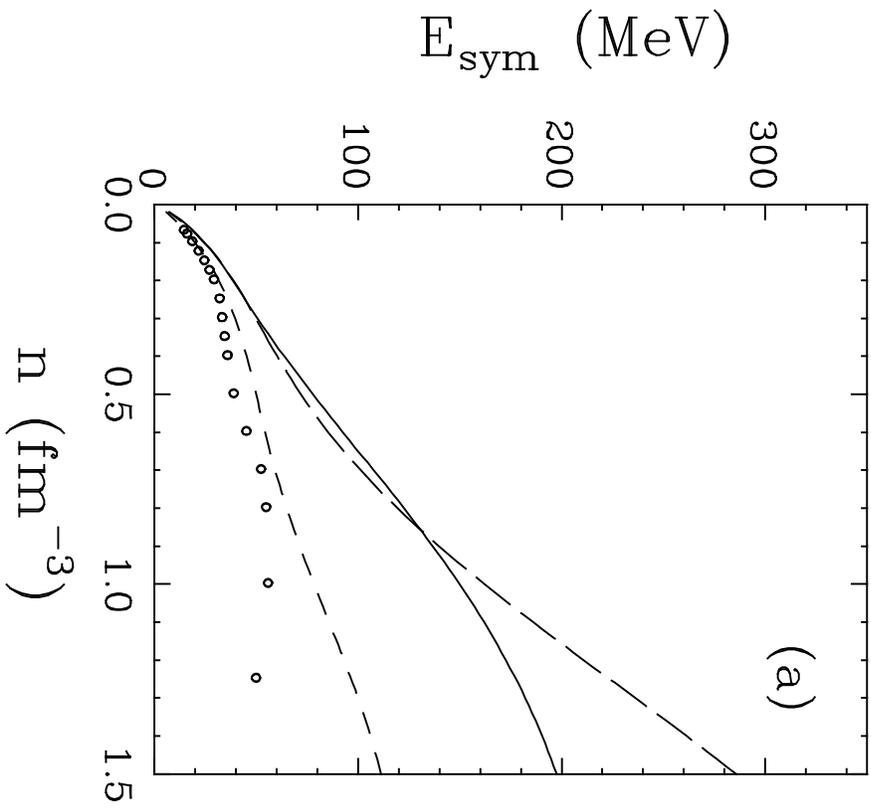
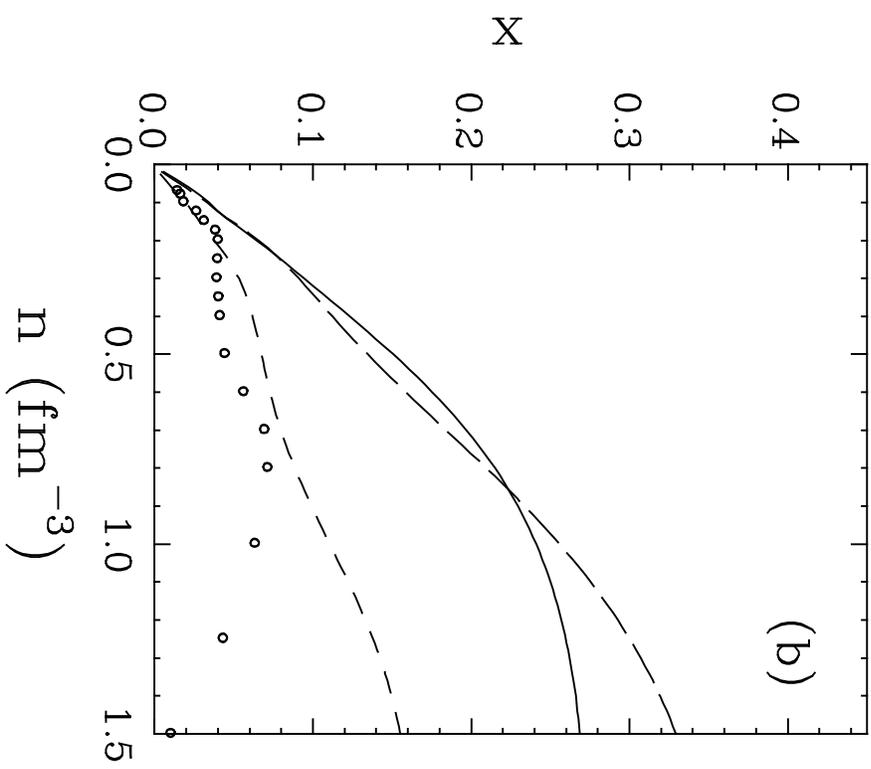
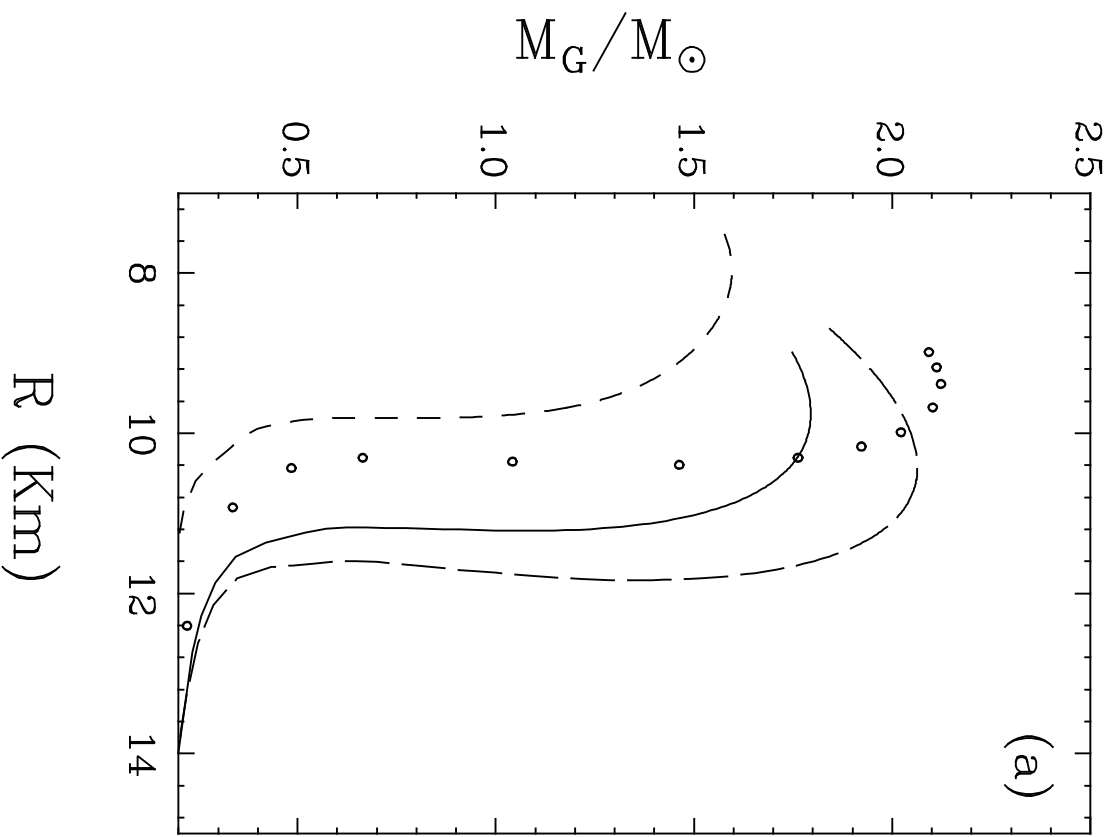
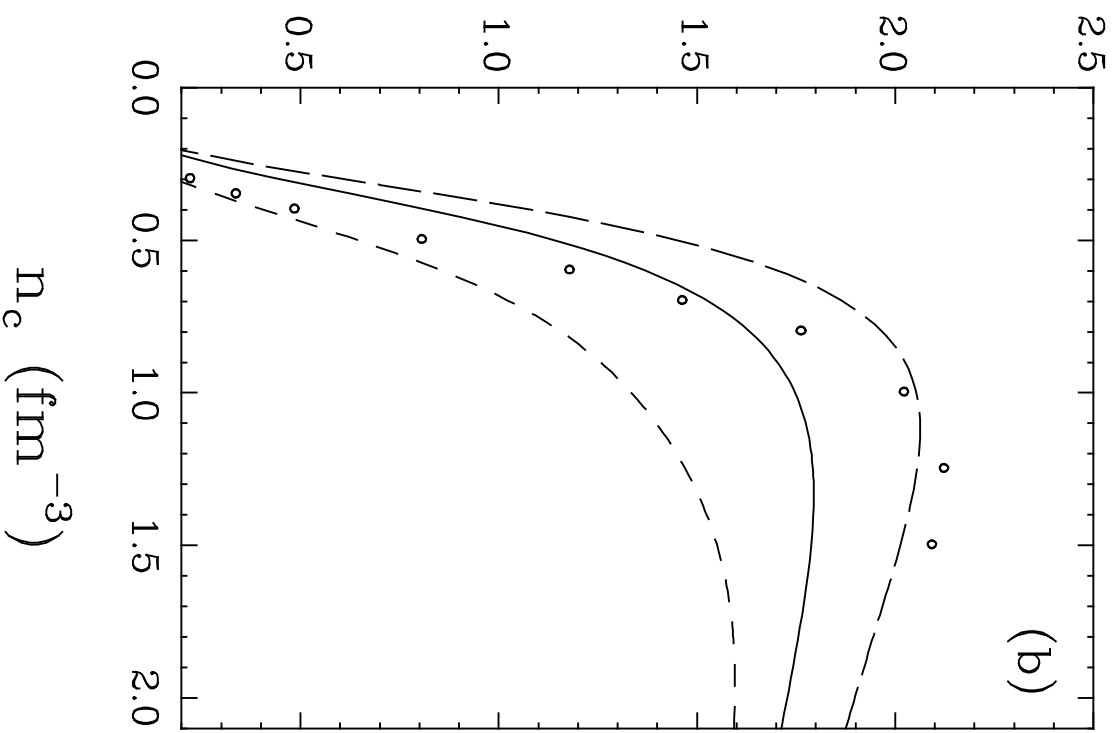
\end{document}